\documentclass[prb,aps,amsmath,amssymb,nofootinbib,epsfig,showpacs,twocolumn,eqsecnum,groupedaddress]{revtex4-1}

\usepackage{amsmath,amssymb,amsfonts,graphicx,multirow,color,bm,textcomp,mathtools}
\usepackage{paralist}
\usepackage{srcltx}
\usepackage[all,cmtip]{xy}
\usepackage{mathrsfs}
\usepackage{srcltx,soul,subfigure}
\usepackage{threeparttable,dsfont,bbm}

\newcommand{\bra}[1]{\langle #1 |}
\newcommand{\ket}[1]{| #1 \rangle}

\newcommand{\bee}{\begin{eqnarray}}
\newcommand{\ee}{\end{eqnarray}}
\newcommand{\bma}{\begin{pmatrix}}
\newcommand{\ema}{\end{pmatrix}}
\newcommand{\balig}{\begin{align}}
\newcommand{\ealig}{\end{align}}

\newcommand{\bZ}{\mathbb{Z}}
\newcommand{\ba}{\begin{align}}
\newcommand{\ea}{\end{align}}

\newcommand{\ignore}[1]{}

\newcommand{\six}{\sigma_x}
\newcommand{\siy}{\sigma_y}
\newcommand{\siz}{\sigma_z}

\newcommand{\bk}{{\bf k}}

\usepackage{dcolumn}
\newcolumntype{C}[1]{>{\centering\let\newline\\\arraybackslash\hspace{0pt}}m{#1}}

\begin{document}

\title{Induced spectral gap and pairing correlations from superconducting proximity effect}

\author{Ching-Kai Chiu}
%\email{chiu7@phas.ubc.ca}

\author{William S. Cole}
%\affiliation{Condensed Matter Theory Center, Department of Physics, University of Maryland, College Park, MD 20742, USA}

\author{S. Das Sarma}
%\affiliation{Condensed Matter Theory Center, Department of Physics, University of Maryland, College Park, MD 20742, USA}
\affiliation{Condensed Matter Theory Center and Joint Quantum Institute and Maryland Q Station, Department of Physics, University of Maryland, College Park, MD 20742, USA}

\begin{abstract}

We theoretically consider superconducting proximity effect, using the Bogoliubov-de Gennes (BdG) theory, in heterostructure sandwich-type geometries involving a normal s-wave superconductor and a non-superconducting material with the proximity effect being driven by Cooper pairs tunneling from the superconducting slab to the non-superconducting slab. Applications of the superconducting proximity effect may rely on an induced spectral gap or induced pairing correlations without any spectral gap. We clarify that in a non-superconducting material the induced spectral gap and pairing correlations are independent physical quantities arising from the proximity effect. This is a crucial issue in proposals to create topological superconductivity through the proximity effect. Heterostructures of 3D topological insulator (TI) slabs on conventional s-wave superconductor (SC) substrates provide a platform, with proximity-induced topological superconductivity expected to be observed on the ``naked" top surface of a thin TI slab. We theoretically study the induced superconducting gap on this naked surface. In addition, we compare against the induced spectral gap in heterostructures of SC with a normal metal or a semiconductor with strong spin orbit coupling and a Zeeman splitting potential (another promising platform for topological superconductivity). We find that for any model for the non-SC metal (including metallic TI) the induced spectral gap on the naked surface decays as $L^{-3}$ as the thickness ($L$) of the non-SC slab is increased in contrast to the slower $1/L$ decay of the pairing correlations. Our distinction between proximity-induced spectral gap (with its faster spatial decay) and pairing correlation (with its slower spatial decay) has important implications for the currently active search for topological superconductivity and Majorana fermions in various superconducting heterostructures.
%Adjusting the Fermi level above the bulk gap (which is the case in experiments). {\clr semiconductor, normal metal discussion. $L^{-3}$ decay, triplet pairing induced.   
%}
\end{abstract}

\date{\today}

\maketitle

%%%%%%%%%%%%%
%%%%%%%%%%%%%%%%%%

\section{Introduction}

When a superconductor (SC) is in electrical contact with a non-SC metal, Cooper pairs from the SC ``leak" into the metal. This proximity effect (PE) is a well-known phenomenon, which was intensely studied in the 1960s\cite{Park_SC_DeGennes,PhysRev.132.1576,clarke:jpa-00213516}. Recently the PE has seen renewed interest as proximity-induced superconductivity provides a promising path to topological superconductivity (TSC)\cite{Kitaev2001,review_TIb}. The induced spectral gap from the PE (which is distinct from pairing correlations arising from the leaking Cooper pairs) is a crucial ingredient for realizing TSC, and indeed induced gaps have been clearly observed in semiconductor nanowires\cite{2016arXiv160304069Z,ChangW.:2015aa} and topological insulator (TI) surfaces\cite{wang_Xue_science_12,SC_Proximity_Jia,Xu_TI_SC}, two promising platforms for TSC.

The theoretical proposal for proximity-induced 2D TSC in a heterostructure of strong 3D TI with a conventional s-wave SC was provided by Fu and Kane\cite{FuKane_SC_STI}. Since the TI surface possesses a single Dirac cone\cite{Roy2009_3D,Moore2007uq,Fu2007uq}, inducing superconductivity at this surface produces an effective $p\pm ip$ TSC.  As the chemical potential is located near the neutral point of the surface Dirac cone\cite{Chiu:2011fk,Chiu_SC_TI_extended,Hosur:2011uq}, when applying a magnetic field, Majorana zero-energy modes (MZMs) are expected to appear in vortices on the ``naked" (top) TI surface as well as at the interface of SC and TI\cite{PhysRevB.81.241310}. At present, there is significant experimental interest in the induced gap on this naked surface of thin TI slabs grown on SC substrates, as it is much more accessible than the buried TI/SC interface. Such SC/TI heterostructures are being actively studied experimentally as possible platforms for realizing Majorana fermions, and hence, topological quantum computation\cite{Sarma:2015aa,RMP_braiding,Kitaev2001,Ivanov_braiding}.

Bi$_2$Se$_3$/NbSe$_2$ and Bi$_2$Te$_3$/NbSe$_2$ heterostructures have been realized experimentally. The induced spectral gap on the naked TI surface was measured varying the TI slab thickness. It was seen that the gap almost vanishes when the thickness exceeds $10$nm\cite{SC_Proximity_Jia,Xu_TI_SC}. However, long-distance induced superconductivity was reported by observing the PE-induced zero resistance state in Bi$_2$Se$_3$ over a distance of $\sim 1\mu$m\cite{Proximity_Josephson}. It turns out that two independent induced physical quantities from the PE have been measured. The long distance of the zero resistance state stems from the slow decay of pairing correlations as the thickness is increased, whereas the spectral gap exhibits a rapid decay.  Throughout this paper, we make the crucial distinction between the mere existence of a pairing-induced supercurrent (i.e. just a zero-resistance state) and the existence of  a spectral gap (which entails necessarily a zero-resistance state, but more) in PE induced superconductivity.  This distinction has not always been clearly made in the PE literature, and typically, discussions of PE have conflated these two phenomena as one and the same (as it would be for real superconductivity\cite{two_SC_proximity}).  We establish in the current work a clear and sharp distinction between PE induced pairing correlation and spectral gap.

Tentative evidence for MZMs in SC/TI platform has been reported from the observation of a zero bias peak\cite{Majorana_Fu_Kane_Jia} and spin-selective Andreev reflection\cite{2016arXiv160302549S} in scanning tunneling microscopy/spectroscopy on a vortex of  Bi$_2$Te$_3$/NbSe$_2$ heterostructure. An alternative proposal by Sau et al.\ takes a thin slab of a semiconductor (SM) with strong spin-orbital coupling sandwiched by an s-wave SC and a ferromagnetic insulator. Through the mechanisms of SC and magnetic proximity effects, the SM slab likewise becomes an effective $p+ip$ superconductor\cite{Sau_semiconductor_heterostructures,PhysRevB.82.214509,PhysRevB.82.094522}. Similarly, a vortex in this heterostructure is able to host a MZM. In this manuscript, we consider three models for non-SC slabs (normal metal (NM), TI, and SM) to investigate the spectral gap and pairing correlations induced through the PE. In our work, we restrict to a theory of PE in 2D sandwich structures involving  SC and non-SC slabs, but our qualitative conclusions apply equally well to heterostructures containing non-SC nanowires on SC substrates, which have been of much interest lately in the context of Majorana fermions\cite{Mourik_zero_bias,Roman_SC_semi,Gil_Majorana_wire,Albrecht:2016aa}. This is because the SC-nanowire heterostructure can be thought of as essentially being a quasi-2D system where one lateral dimension of the 2D non-SC slab has been shrunk down to create the 1D nanowire.

The spectral gap and pairing correlations induced by PE have been theoretically studied separately in the literature\cite{Park_SC_DeGennes,PhysRev.132.1576,clarke:jpa-00213516,deGennesbook}. Comparison of these two different physical quantities as well as explanation of the sharp distinction between them are lacking, however. In this manuscript, we clarify that at zero temperature the spectral gap and pairing correlations induced by the superconducting PE in non-SC materials exhibit different behaviors indicating subtle differences in their physical origin.
In conventional BCS superconductors, the spectral gap is given by the \emph{order parameter}, which is directly connected to the pairing correlation function by a non-zero attractive electron-phonon interaction.
For non-SC materials, with no intrinsic electron-phonon interaction, the order parameter is strictly vanishing, and the corresponding induced quantities need not be related; indeed the two are completely distinct, and it is possible to create PE-induced supercurrent without having a spectral gap.
The induced gap plays an important role in determining the robustness of MZMs on the surface of a 3D topological insulator (or the 1D or 2D semiconductor system), while the induced pairing, which has been theoretically studied\cite{SC_Proximity_Kim}, is related to the Josephson current and does not affect the presence of MZMs. For example, the observation of induced supercurrent on a quantum spin Hall edge\cite{Induced_SC_QSH,PhysRevB.90.224517,Edge_SC_Leo} indicates the presence of induced pairing, but no spectral gap exists in these experiments implying that MZMs cannot be created here in spite of the PE-induced flow of supercurrent in the non-SC material.  Thus, the experiments observing only supercurrents (and thus manifesting only pairing correlations) are qualitatively distinct from experiments observing spectral gaps in the PE literature.  We emphasize that this distinction, which does not exist in a real SC, arises only in the context of PE in non-SC materials by virtue of the fact that the non-SC system has no intrinsic (e.g. electron-phonon coupling induced) attractive interaction forcing the spectral gap and pairing correlation to be directly connected with each other.

%When the induced gap on the naked TI surface is large enough, MZMs are able to be localized and observed.
%At present, there is significant experimental interest in the induced gap on the naked (top) surface of thin TI slabs grown on SC substrates.
%When the induced gap on the naked surface (as opposed to the buried interface) is large enough, MZMs are able to be localized and observed on the vortices produced by an external magnetic field \cite{Chiu:2011fk}, and due to their presence such a heterostructure provides a promising platform for experimental detection of MZMs.

%{\clr  mention paper structure }

The remainder of this paper is organized as follows. In Sec.~\ref{model} we describe three model Hamiltonians for NM/SC, TI/SC, and SM/SC heterostructures. 
Sec.~\ref{gap} is devoted to the study of the induced spectral gap on the naked surface of the non-SC slabs.
We show in Sec.~\ref{pairing} the spatial distribution of induced pairing correlations in the heterostructures.
Finally, in Sec.~\ref{summary} we conclude the paper and give an outlook on future research.

%------------------------------------------------------------------------------%

\section{Model} \label{model}

%{\clr add semiconductor with strong spin orbit}

\begin{figure}[t!]
\begin{center}
\includegraphics[clip,width=0.98\columnwidth]{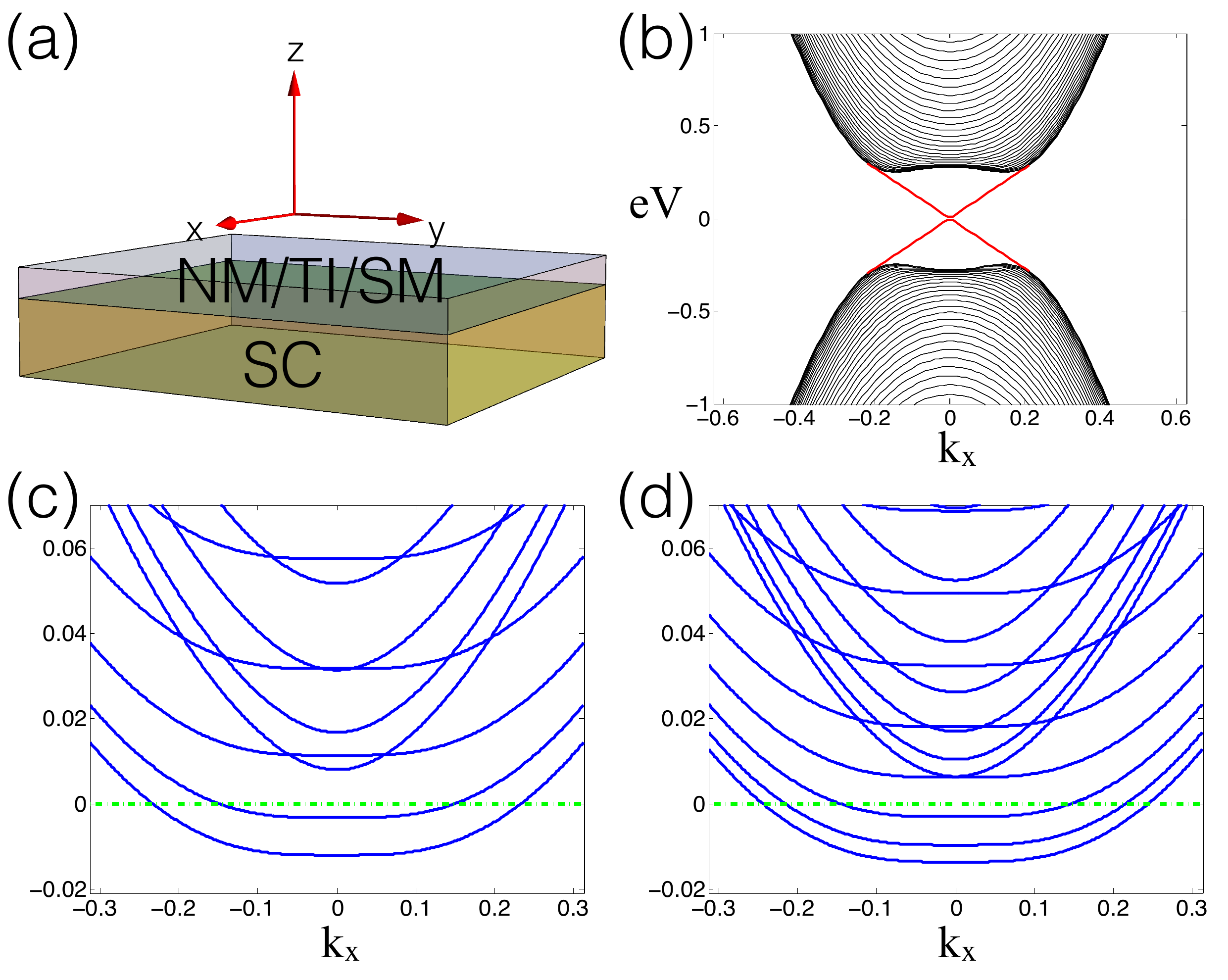}
\end{center}
\caption{(a) The heterostructure of non-SC slab on a superconductor substrate. (b) The energy spectrum of the TI Bi$_2$Se$_3$. Red and black indicate surface and bulk spectra respectively. (c) and (d) show spectra for the SM model with even and odd numbers of energy bands crossing at the Fermi level for $L_{\rm{SM}}=40$ and $60$ respectively. 
}   
\label{heterostructure_TI_spectrum} 
\end{figure}

The Fu-Kane model\cite{FuKane_SC_STI} describes a strong topological insulator (TI) surface state proximity-coupled to a conventional s-wave superconductor (SC).
In recent experiments designed to realize the Fu-Kane model, the TI samples were deliberately made metallic (rather than insulating) in the bulk in order to enhance the free carrier density and hence the induced superconductivity.
The difference between these \emph{metallic} topological insulators and normal metals is the presence of stable gapless states at the TI surfaces in addition to the metellic bulk.
To determine whether the superconducting proximity effect is independent of material details, we numerically study NM, TI, and SM slabs as non-SC materials (which have no intrinsic attractive interaction).
The proximity effect can be realized in a heterostructure of a non-SC slab and an s-wave superconductor as illustrated in fig.~\ref{heterostructure_TI_spectrum}~(a).
In this section, we give the heterostructure Hamiltonians in the form of BdG tight-binding models for these three distinct non-SC materials.
The induced superconducting gap on the naked (top) surface of the non-SC material and induced pairing correlations can be obtained by numerically solving the corresponding BdG eigenvalue problem.

\subsection{Normal Metal (NM)}

We start with a toy model of a conventional NM ($0<z\leq L_{\rm{NM}}$) slab on top of an s-wave superconductor ($-L_{\rm{SC}}< z \leq 0$) as in fig.~\ref{heterostructure_TI_spectrum}~(a).
With periodic boundary conditions in the $x, y$ directions the BdG Hamiltonian is given by
\begin{subequations}\label{HNM}
\bee
\hat{H}^{\rm{NM}}=\hat{H}_{\rm{NM+SC}} + \hat{H}_t,
\ee
where
\begin{align}
\hat{H}^{\rm{NM}} =& -(c(k_x,k_y)+ \mu)\sum_{-L_{\rm{SC}} < z \leq L_{\rm{NM}}} \Theta^\dagger_z \sigma_z \Theta_z \nonumber \\
- &\frac{1}{2} \sum_{-L_{\rm{SC}} < z < L_{\rm{NM}}}^{z\neq 0} (\Theta^\dagger_{z+1}  \sigma_z \Theta_z + h.c. ) \nonumber \\
 + & \Delta \sum_{-L_{\rm{SC}} < z \leq 0 }\Theta^\dagger_z \sigma_x \Theta_z,  \\
 \hat{H}_t=& t \Theta^\dagger_1 \sigma_z \Theta_0 +h.c., \label{coupling_NM}
\end{align}\end{subequations}
and $c(k_x,k_y)=\cos k_x+ \cos k_y$. Fermion annihilation and creation operators are grouped into the Nambu vector $\Theta_z=(c_{z\uparrow}^{}\ c_{z\downarrow}^\dagger)^T$.
Spin $SU(2)$ symmetry is preserved, resulting in an identical copy of this Hamiltonian in the opposite spin basis, so that this Hamiltonian $\hat{H}^{\rm{NM}}$ is sufficient to describe the entire system.
In this toy model, the parameters are in units of the hopping matrix element, which is taken to be identical for SC and NM for the sake of simplicity in notations (without any loss of generality), and the lattice constant is unity (i.e. the unit of length) so that $z$ is an integer.
Non-zero $\Delta$ for $-L_{\rm{SC}} < z \leq 0$ describes the intrinsic superconducting gap in the SC inducing the PE.
For our numerical study, we choose specific parameters $L_{\rm{SC}}=40, \Delta=0.1$ and $\mu=-2- \cos 0.1$ (i.e., slightly below the band bottom for a very thin slab), and fix $k_y=0$.
The strength of the hopping between the SC and NM layers $t$ is chosen to be $-1/2$.

\subsection{Topological Insulator (TI)}

For the topological insulator model, the BdG Hamiltonian can be written as
\begin{subequations} \label{HFK}
\bee
\hat{H}^{\rm{TI}}=\hat{H}_{\rm{TI}}+\hat{H}_{\rm{SC}}+\hat{H}_T
\ee
where 
\begin{small}
\begin{align}
\hat{H}_{\rm{TI}}=&\sum_{0<z \leq L_{\rm{TI}}} \Pi_z^\dagger \Big \{ \big [ M -2B_1 - 2B_2 (2- c(k_x,k_y)) \big ] \tau_z \sigma_z s_0 \nonumber \\
+& A_2 \sin k_x  \tau_0 \sigma_x s_x + A_2 \sin k_y \tau_z \sigma_x s_y - \mu \tau_z \sigma_0 s_0 \Big \} \Pi_z \nonumber \\
+&\sum_{0<z < L_{\rm{TI}}} \big [ \Pi_{z+a}^\dagger (B_1\tau_z \sigma_z s_0 - \frac{iA_1}{2}\tau_0 \sigma_x s_z)\Pi_z + h.c. \big ], \label{TI_FK} \\
\hat{H}_{\rm{SC}}=&\sum_{-L_{\rm{SC}}<z \leq  0} \Pi^\dagger_z \big [ M({\bf k})\tau_z+\Delta\tau_y \sigma_0 s_y \big ] \Pi_z  \nonumber \\
+&\sum_{-L_{\rm{SC}}< z < 0} \big [ \Pi_{z+a}^\dagger D_1 \tau_z \sigma_0 s_0 \Pi_z +h.c. \big ], \\
\hat{H}_T=&T\Pi^\dagger_1 \tau_z \sigma_0 s_0 \Pi_0 + h.c.,
\end{align}
\end{small}
\end{subequations}
where $M({\bf k})=2 (M_{\rm{SC}}-D_1 (\cos k_x + \cos k_y))$.
Fermion annihilation and creation operators are grouped into a much larger Nambu vector $\Pi=(b_{\uparrow}\ b_{\downarrow}\ d_{\uparrow}\ d_{\downarrow}\ b_{\uparrow}^\dagger\ b_{\downarrow}^\dagger\ d_{\uparrow}^\dagger\ d_{\downarrow}^\dagger)^T$, where $b, d$ describe different orbitals.
In the TI region $0< z \leq L_{\rm{TI}}$, the presence of spin-orbital coupling breaks the spin $SU(2)$ symmetry, so the Hamiltonian, in contrast to the NM model, cannot be written in the half spin basis.
In the SC region $-L_{\rm{SC}} < z \leq 0$, the spin and orbital textures vanish leaving a conventional s-wave superconductor.
The lattice constant is chosen to be $a=3\AA$ for both TI and SC regions so that $z\in \bZ \cdot 3\AA$ and the real lattice momentum $k_i'= k_i/a$.
To make reasonable comparisons with the data from recent experiments, we adopt realistic physical parameters for Bi$_2$Se$_3$ as the TI\cite{Zhang:2009aa} and NbSe$_2$ as the SC with gap $\Delta=1$~meV. 
The remaining parameters adapted from [\onlinecite{Zhang:2009aa}] are in units of eV: $M_{\rm{SC}}=0.28$, $A_1=2.2/a$, $A_2=4.1/a$, $B_1=10/a^2$, $B_2=56.5/a^2$, $D_1=7.6/a^2$, $M_{\rm{SC}}=2.96D_1$.
We choose $L_{\rm{SC}}=45~\AA$, fix $k_y=0$, and assume $T=2$eV is the strong coupling between TI and SC at the interface.
As shown in fig.~\ref{heterostructure_TI_spectrum}~(b) the TI gap is $0.28$~eV.
The chemical potential $\mu$ is adjusted to consider both bulk metallic and insulating values for the TI.

\begin{figure*}[htp!]
\begin{center}
\includegraphics[clip,width=1.5\columnwidth]{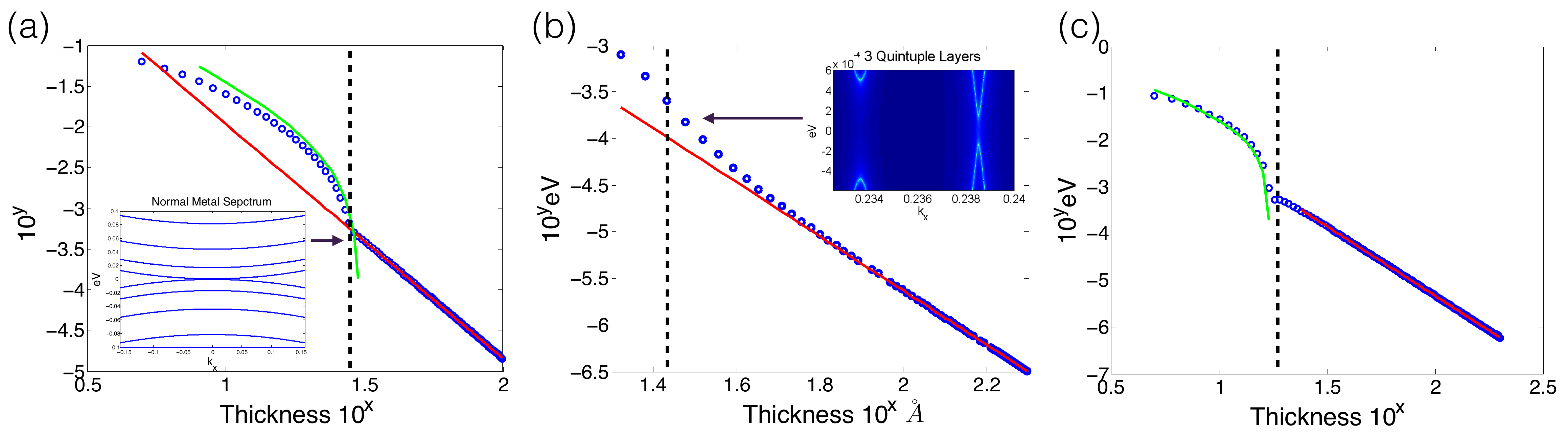}
\end{center}
\caption{Panels (a) show the smallest gap decays for the toy model of NM as a function of $L_{\rm{NM}}$ in log-log scale, respectively. Panels (b) and (c) show the same for Bi$_2$Se$_3$ and the semiconductor respectively. The red fitting lines show that for all of the models the exponent of the power law decay is close to $-3$. The transition from exponential to power law decay (dashed lines) occurs when the bulk bands of the non-superconducting material start to touch at zero energy for some $L$.  The inset of (a) indicates the bulk band touching of the normal metal at $L_{\rm{NM}}=29$, which is the transition point of the decay behavior. The inset of (b) shows the density of states on the naked surface of the TI for 3 quintuple layers (one quintuple layer is roughly $10~\AA$). On the naked surface, the gap is about $0.15$~meV.
}   
\label{Gap_decay} 
\end{figure*}

\subsection{Semiconductor with spin orbit coupling (SM)}
	Finally, we consider a semiconductor (SM) slab with Rashba spin orbital coupling sandwiched between an s-wave superconductor and a ferromagnetic insulator\cite{Sau_semiconductor_heterostructures}. The role of the ferromagnetic insulator is to open a magnetic gap at the $\Gamma$ point through a ferromagnetic proximity effect. Since our focus is mainly on the SC proximity effect, to simplify the problem we replace the ferromagnetic insulator with a uniform Zeeman field in the SM (without worrying about any orbital magnetic field effects, i.e., we simply assume a spin-spin splitting in the SM). The effective Hamiltonian for this model is described by SC and SM parts 
\begin{subequations}	
\bee
\hat{H}^{\rm{SM}}=\hat{H}_{\rm{SC}}+\hat{H}_{\rm{SM}}+\hat{H}_{t'} \label{HSM}
\ee	
where
\begin{small}
\begin{align}
\hat{H}_{\rm{SC}}=& \sum_{-L_{\rm{SC}}<z \leq 0} C_z^\dagger \big [ \big ( -c(k_x,k_y)+ 3 -\mu_{\rm{SC}} \big )\tau_z + \Delta \tau_y \siy \big ] C_z \nonumber \\
&- \frac{1}{2}\sum_{-L_{\rm{SC}}<z < 0} (C^\dagger_z \tau_z C_{z+1}+h.c. ),  \\
\hat{H}_{\rm{SM}}=& \sum_{0<z \leq L_{\rm{SM}}} C_z^\dagger \big [ \big ( -c(k_x,k_y)+ 3 -\mu_{\rm{SC}} \big )\tau_z \nonumber 
 \\ & + V(\sin k_y \tau_0 \six - \sin k_x \tau_z \siy ) + F\tau_z \siz\big ] C_z \nonumber \\
&- \frac{1}{2}\sum_{1<z < L_{\rm{SM}}} (C^\dagger_z \tau_z C_{z+1}+h.c. ),  \\
\hat{H}_{t'}=&t'C^\dagger_1 \tau_z C_0 +h.c. ,
\end{align} 
\end{small}
\end{subequations}
\noindent and $z$ is an integer. We choose the pair potential ($\Delta=0.1$) and chemical potential ($\mu_{\rm{SC}}=1$) in the SC, the thickness ($L_{\rm{SC}}=25$) of the SC, the strength ($V=0.1$) of spin orbital coupling, the coupling ($t'=1/2$) between SC and SM, and the Zeeman field ($F=-0.01$) in the $z$ direction. The chemical potential in the semiconductor is tuned to $\mu_{\rm{SM}}=0.005$ to be located at the gap opened by the Zeeman field and $k_y=0$ to simplify the problem. %One of the main goals is to investigate the induced SC gap in the SM as the SM thickness is increased.
A MZM is expected to appear in a vortex in this heterostructure. The criterion for the presence of the MZM is that an odd number of the SM bands cross the Fermi level. However, as the SM thickness ($L_{\rm{SM}}$) is increased, the number of crossing bands increases since the number of discrete momenta in the $z$ direction is proportional to $L_{\rm{SM}}$. As shown in fig.~\ref{heterostructure_TI_spectrum}~(c,d), if the chemical potential is fixed, the topological and trivial phases alternate as $L_{\rm{SM}}$ is varied since the system goes through odd and even number of bands crossing the chemical potential.

%\begin{figure}[t!]
%\begin{center}
%\includegraphics[clip,width=0.98\columnwidth]{fig8}
%\end{center}
%  \caption{  (a) Energy spectrum of the SM as $k_z=0$. The chemical potential is located the gap opened by the Zeeman field. (c-d) show different numbers of energy bands crossing the Fermi level. 
%}   
%\label{spectrum} 
%\end{figure}

%------------------------------------------------------------------------------%

\section{Spectral gap} \label{gap}

To find the induced superconducting gap ($\Delta_{I}$) on the naked surface, it is reasonable to assume that the smallest energy gap from the entire spectrum of the heterostructure should occur on the naked surface.
However, this energy gap is not necessarily the induced superconducting gap.
To confirm the gap origin, we can simply turn off the coupling between SC and non-SC ($t,\ T,\ t'=0$) and compare the spectra.
From exact diagonalization, we obtain the smallest energy gap of the NM, TI, and SM heterostructures for different thickness $L_{\rm{NM}}$, $L_{\rm{TI}}$, and $L_{\rm{SM}}$ as shown in fig.~\ref{Gap_decay}.
To simplify the problem, we restrict to $k_y=0$ and find the smallest gap over all $k_x$. For the TI heterostructure, we first choose a chemical potential $\mu=0.4$~eV, which is larger than the TI gap $0.28$~eV, to enhance the induced gap.

\subsection{Metal}

%{\clr Universal Properties}
\begin{figure*}[t!]
\begin{center} 
\includegraphics[clip,width=1.5\columnwidth]{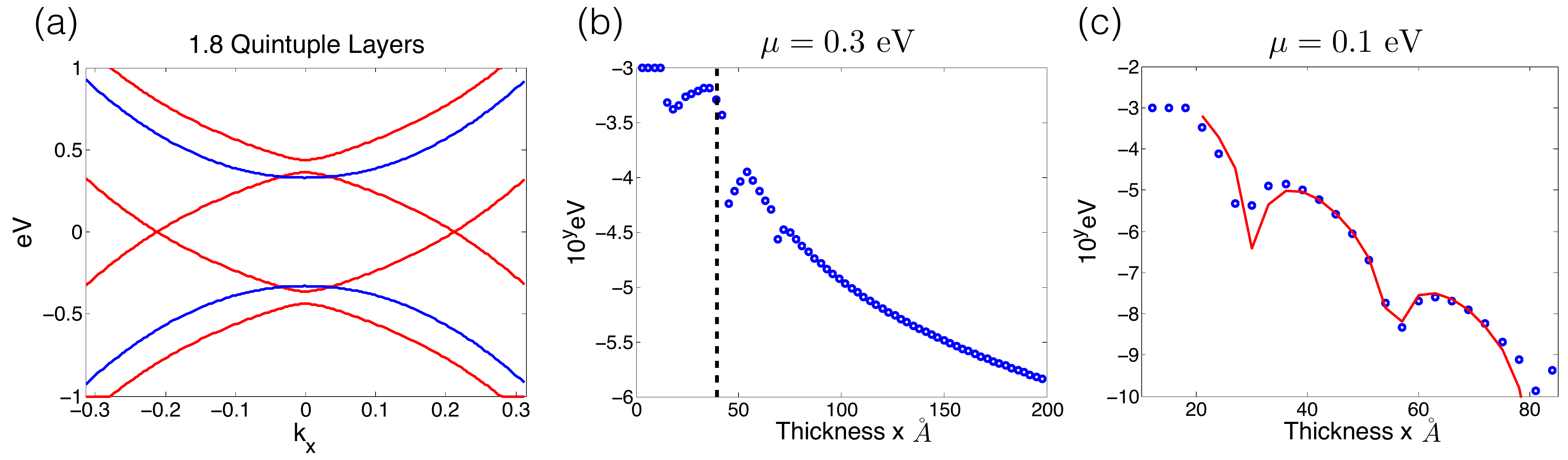}
\end{center}
\caption{In (a) we show the bulk (blue) and surface (red) spectrum of the TI slab at $\mu=0.4$~eV with $L_{\rm{TI}}=18~\AA$. Although the finite size effect causes the bulk and surface band gap, by adjusting the chemical potential the surface states are gapless near zero energy. (b) and (c) show the smallest gap (blue) of the TI heterostructure at $\mu=0.3$ and $0.1$~eV respectively. The dashed line indicates the transition point between bulk insulator and metal and the red line in (c), which is from the portion of the naked surface state at zero energy leaking to SC region (see appendix \ref{surface derivation}), has a great fitting with the numerical result. 
}   
\label{small_L} 
\end{figure*}

These three different models provide several apparently universal results.
First, the gap decay behavior shows a transition (dashed lines in fig.~\ref{Gap_decay}) at a certain thickness ($L_{\rm{NM/TI/SM}}$) of the non-SC slabs.
This can be explained by a transition from insulator to metal. 
That is, due to the finite size effect, all of the models with small $L$ possess a bulk gap due to the simple size quantization effect. The reason for emphasizing this finite size effect is that the TI slab in experiments might be too thin to be metallic in the presence of the trivial size quantization effect. 
When the thickness is increased, the bulk gap is closed by the band touching (e.g. the inset of fig.~\ref{Gap_decay}~(a)).
On the other hand, for the NM and SM heterostructures the transition appears sharp, while there is a much smoother crossover behavior for the TI heterostructure.
The reason is the presence of the gapless surface states in TI for small $L_{\rm{TI}}$, which will be discussed below in detail.
Second, when the thickness is greater than the transition point, the induced gap decay rate is universal in all three systems,
\bee	
\Delta_{I} \propto L^{-3}. 
\ee		
In this regime, the universal decay rate stems from the zero-energy wavefunction of the non-SC side ``leaking" into the SC.

This universal spectral gap decay rate $L^{-3}$, which is a surprising result when compared with the well known $L^{-1}$ decay of the superconducting \emph{pairing} predicted by Deutscher and de Gennes~\cite{Park_SC_DeGennes}, deserves a detailed physical interpretation.
We use the NM heterostructure to explain this universal property, though the metallic TI and SM can also be explained the same way. First, consider the situation with the coupling (\ref{coupling_NM}) between the NM and SC turned off. The states in the $z$ direction can be described by the wavefunctions in the 1D infinite square well  
\bee
\phi_z^n= \sin (k_z^n z ), \label{well}
\ee
where $k_z^n=\pi n/L$ and $n$ is a positive integer. The energy spectrum of the NM in momentum space (again, having taken $k_y = 0$) is given by 
\bee
E_{\pm}^n = \mp (\cos k_x^n + \cos k_z^n +1+\mu)
\ee		
Now turning on the coupling between the NM and SC, we expect that the NM states do not dramatically change but small portions of the NM states leak to the SC region. 
Based on our simulations, the state that acquires the smallest induced gap can be described by $\phi_z^1= \sin (\pi z/L )$. 
%as the bulk states are sensitive to the boundary. 
Since this wavefunction has zero energy in the absence of coupling to the SC, the momentum $\phi_{K_x}$ in the $x$ is given by 
\bee
K_x\sim \cos^{-1}(-\cos(\frac{\pi}{L})-1-\mu) .
\ee	
The leaking portion $\phi_{K_x}^{\rm{leak}}$, which represents the support of this state on the SC side of the interface, can create an effective gap in the SC region 
\begin{align}
 \Delta_{I} &= \bra{\phi_{K_x\downarrow\rm{h}}^{\rm{leak}}}H_{\rm{SC}}\ket{\phi_{K_x\uparrow\rm{p}}^{\rm{leak}}} \sim \Delta|\phi_{K_x}^{\rm{leak}}|^{2},
\end{align}
where $H_{\rm{SC}}$ is the Hamiltonian of the superconductor with gap $\Delta$, and $\downarrow\rm{h}$ and $\uparrow\rm{p}$ indicate the spin down hole and spin up particle amplitudes respectively.

For the NM heterostructure with several arbitrary parameter choices, numerical computation of $|\phi_{K_x}^{\rm{leak}}|^{2}$ with different $L$ results in 
\bee 
|\phi_{K_x}^{\rm{leak}}|^{2}\propto L^{-3}
\ee
In fact, the leaking portion of the wavefunction can further be related to the wavefunction $\phi_{K_x}^{\rm{boundary}}$ at the edge of the NM.
That is, in the lattice model $\phi_{K_x}^{\rm{boundary}}$ is given by the wavefunction of the pure NM with length $L$ on the end lattice site.
The reason is that when the SC is present, the continuity condition at the interface between SC and NM leads to $\phi_{K_x}^{\rm{boundary}} \sim \phi_{K_x}^{\rm{leak}}$.
Again by computing the wavefunction in the normal metal for several different parameters we confirmed that the same power-law decay is still present,
\bee
|\phi_{K_x}^{\rm{boundary}}|^2 \propto L^{-3}
\ee
Evaluating the gap is thus simplified to the boundary problem for the NM wavefunction. The essential physics can be demonstrated by the wavefunction in the 1D infinite square well extended to the SC region ($-\epsilon<z<L$), where $\epsilon$ is an unknown positive constant close to zero. The normalized wavefunction, mainly in NM region, along the $z$ direction is approximately described by  
\bee
\phi_{K_x}= \sqrt{\frac{2}{L+\epsilon }}\sin(\frac{\pi (z +\epsilon) }{L+\epsilon}),
\ee
The wavelength is slightly extended from $L$ to $L+\epsilon$ since the small portion of $\phi_{K_x}$ leaks into the SC region. At the boundary of SC and NM, $\phi_{K_x}^{\rm{boundary}}$ is determined by the value of the wavefunction at $z=0$ and small $\epsilon$, $\phi_{K_x}^{\rm{boundary}} \propto L^{-3/2}$.
Thus, $\Delta_{I} \propto L^{-3}$. 

\subsection{Insulator}

When the thickness $L$ of the non-SC slab is small, the bulk spectrum is gapped.
In this insulator region, the smallest gap decays of the NM(SM) and TI exhibit distinct behaviors.
The smallest gap of the NM(SM) heterostructure (c.f. fig.~\ref{Gap_decay}~(a,c)) monotonically decays whereas the smallest gap (c.f. fig.~\ref{small_L}~(b,c)) of the TI heterostructure decays with an oscillation due to the presence of the gapless surface states.
Since fig.~\ref{Gap_decay}~(b) has a small insulating region of slab thickness $L_{\rm{TI}}$, to extend the insulating region we adjust the chemical potential to $0.3$ and $0.1$~eV, and show the decay of the minimum gap in fig.~\ref{small_L}~(b,c). Due to the bulk gap ($0.28$~eV), the insulator to metal transition never occurs for $\mu=0.1$~eV.

We first discuss the NM(SM) heterostructure.
For small $L$, the smallest gap stems from the finite size effect of the NM(SM) with open boundary condition, rather than the induced superconducting gap.
As shown in fig.~\ref{Gap_decay}, the green line, calculated for a slab of NM with thickness $L$ and no SC coupling, perfectly fits the smallest gap of the NM(SM) heterostructure.
The dashed line for the NM(SM) represents the transition between the gap arising from a finite size effect and the induced superconducting gap.

The smallest gap in the bulk insulating TI ($\mu = 0.1$eV), which exponentially decays with oscillation, always represents the induced superconducting gap.
The reason is that, since the chemical potential is intentionally adjusted off zero, the only other possible gap in the TI is the finite size gap of the surface state which has been excluded (see fig.~\ref{small_L}~(a)). 
The reason for the decay behavior is that the spatial distribution of the gapless states on the naked surface (see Appendix \ref{surface derivation}) also exponentially decays with oscillation.
The portion $\Psi_{\rm{surf}}^{\rm{leak}}$ of the surface state that leaks to the SC region induces the SC gap. Following similar logic to the $L^{-3}$ decay, we have  
\bee
\Delta_{I} \propto |\Psi_{\rm{surf}}^{\rm{leak}} (L)|^2
\ee	
Appendix \ref{surface derivation} shows the analytic solutions of the surface state wavefunction $\Psi(z)$.
We use the zero-energy wavefunction $\Psi(\epsilon)$ of the naked surface states at the interface to calculate the red line $\zeta|\Psi(\epsilon)|^2$ in fig.~\ref{small_L}~(c) (for which $\epsilon, \zeta$ are fitting constants) and find an excellent fit to the smallest gap of the TI heterostructure.

We prove in Appendix \ref{surface derivation} that the decay rate of the surface wavefunction is independent of the chemical potential.
Hence, the chemical potential also does not change the decay length of the induced gap. 
In this model of a Bi$_2$Se$_3$/NbSe$_2$ heterostructure we obtain an effective decay length of $8.75~\AA$ from the gapless surface wavefunctions. This estimated decay length is too short in comparison with measurements of the induced superconducting gap\cite{SC_Proximity_Jia,Xu_TI_SC}.
However, a recent APRES experiment\cite{Xu_TI_SC} shows the length $L_{\rm{TI}}$ for the insulator/metal transition is less than $40~\AA$.
Entering the metal region, which exhibits $L^{-3}$ decay, explains the slower induced gap decay seen in experiments\cite{SC_Proximity_Jia,Xu_TI_SC}. 

%------------------------------------------------------------------------------%

\section{Superconducting pairing}	\label{pairing}

\begin{figure}[thp!]
\begin{center}
\includegraphics[clip,width=0.98\columnwidth]{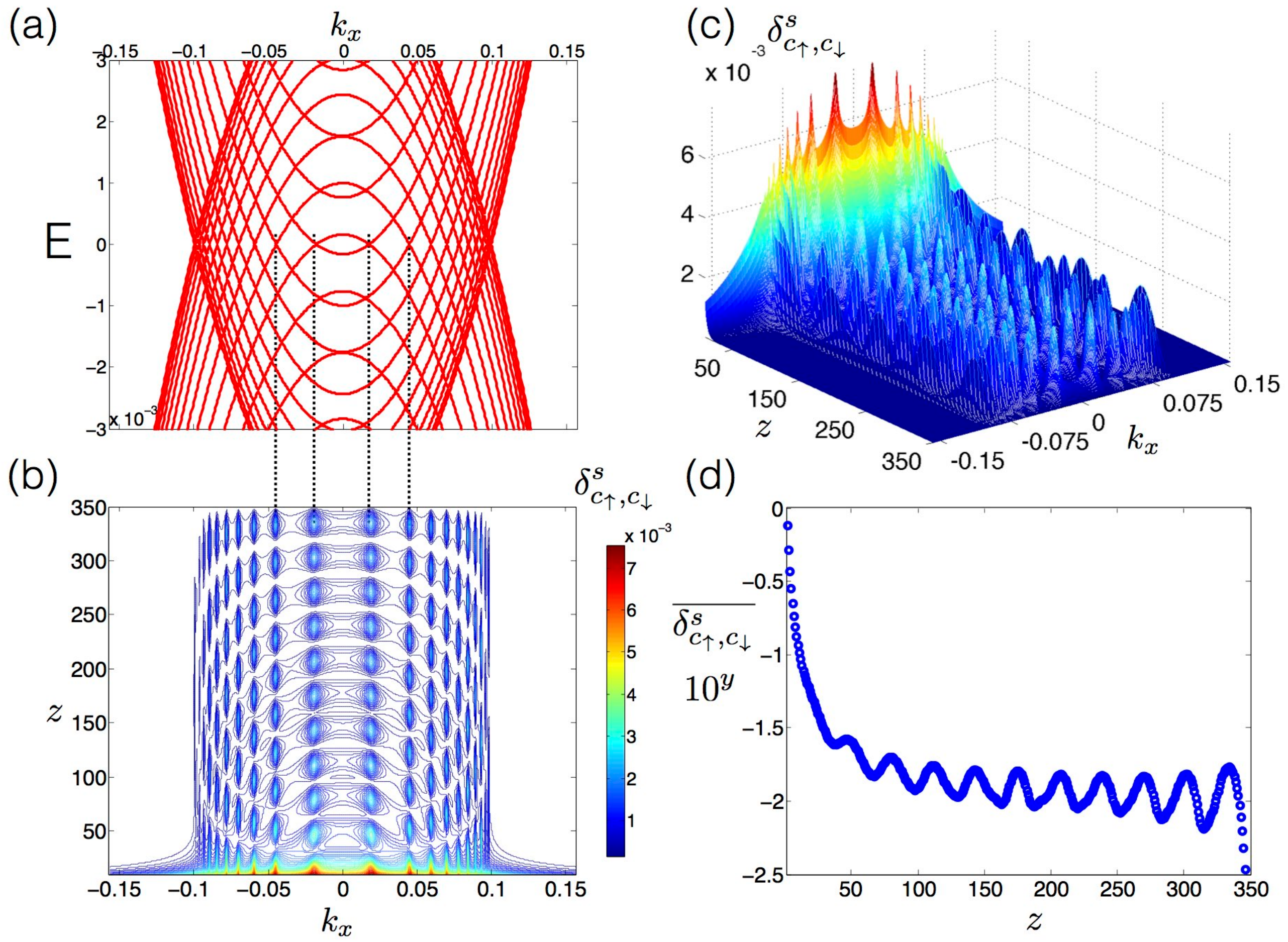}
\end{center}
\caption{Singlet pairing distribution of the NM heterostructure ($L_{\rm{MN}}=350$): (a) the energy spectrum of the NM slab  in the BdG Hamiltonian. Panels (b) and (c) show the pairing  distribution $|\delta^s_{c\uparrow,c\downarrow}|$ as a function of $k_x$ and $z$ in the NM region region ($0<z\leq L_{\rm{NM}}$) and the pairing oscillation in the NM.  (d) shows the distribution of the pairing average $\overline{\delta^s_{c\uparrow,c\downarrow}}$ also decays and oscillates along the $z$ direction. The dashed lines between (a) and (b) are to indicate that non-zero pairing appears where $k_x$ corresponds to zero-energy crossings in the NM spectrum.
}   
\label{NM_pairing} 
\end{figure}

The superconducting proximity effect induces \emph{pairing} in addition to a spectral gap in non-superconducting materials.
Unlike the spectral gap, however, the induced pairing is calculated (in the BdG formalism) from \emph{all} of the occupied states $\ket{\Psi_{E\leq 0}}$ in the heterostructure.
We can quantify the pairing in the generic form 
\begin{subequations}
\begin{align}
\delta^{s/t}_{\alpha,\beta}(\bk)\equiv&\langle \alpha^\dagger_k \beta^\dagger_{-k} \mp \beta^\dagger_{k}  \alpha^\dagger_{-k} \rangle /{2}, \\
\delta^{t}_{\alpha}(\bk)\equiv&\langle \alpha^\dagger_k \alpha^\dagger_{-k} \rangle ,
\end{align}
\end{subequations}
where $\langle \# \rangle \equiv \sum_{E\leq 0} \bra{\Psi_{E}}  \# \ket{\Psi_{E}}$ and $\alpha,\beta$ include both the spin and orbital character of the electron operators. The functions $\delta^t_{\#}(\bk)$ are odd functions in $\bk$, indicating triplet pairing, while $\delta^s_{\#}(\bk)$ are even in $\bk$, indicating singlet pairing. 
%{\clr singlet pairing TI and NM. all pairings TI and SM. }

\subsection{Singlet pairing}

\begin{figure}[t!]
\begin{center}
\includegraphics[clip,width=0.95\columnwidth]{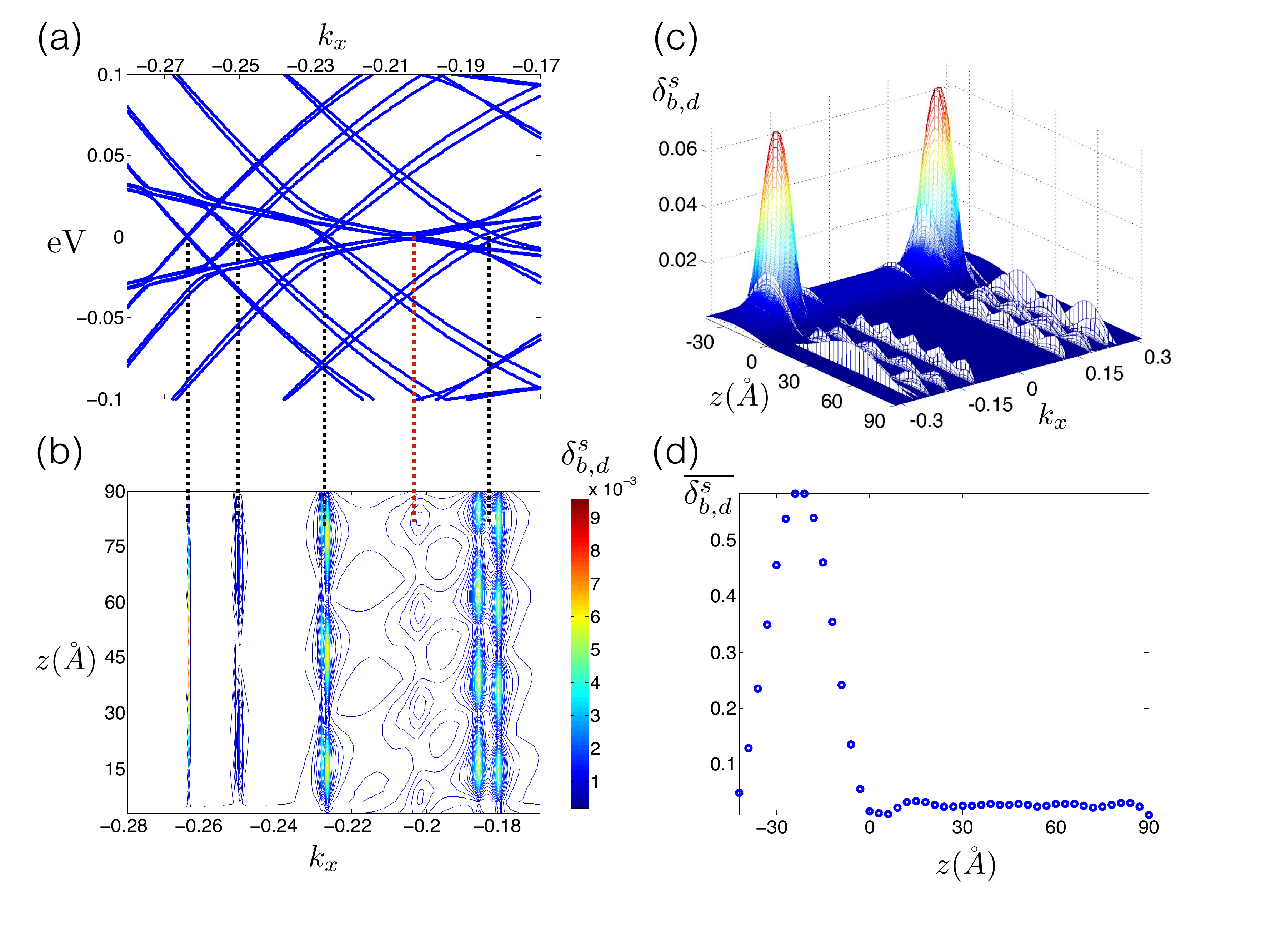}
\end{center}
\caption{Singlet pairing distribution of the TI heterostructure ($L_{\rm{TI}}=90~\AA$). (a) the energy spectrum of the \emph{entire} heterostructure (i.e., both TI and SC bands). Panels (b) and (c) show the pairing distribution $\delta^s_{b,d}$ as a function of $k_x$ and $z$ in the entire heterostructure region ($-L_{\rm{SC}}< z\leq L_{\rm{TI}}$). The pairing almost vanishes at the interface ($z=0$) then oscillates weakly in the TI region.  (d) shows the distribution of the pairing sum $\overline{\delta^s_{b,d}}$ oscillates along the $z$ direction without decay. The black (red) dashed lines between (a) and (b) indicate that non-zero pairing appears where $k_x$ corresponds to zero-energy crossings in the TI(SC) spectrum.
}   
\label{TI_pairing} 
\end{figure}

In the s-wave superconductor slab, the only non-zero pairing present for the NM heterostructure is the singlet $\delta^s_{c_\uparrow,c_\downarrow}$.
Likewise, $\delta^s_{b_\uparrow,b_\downarrow}$ and $\delta^s_{d_\uparrow,d_\downarrow}$ are dominant for the TI heterostructure.
In the previous section, we discussed the induced gap on the naked surfaces of the NM, TI, and SM slabs as their thickness varies.
For the induced pairing, however, our main interest is its spatial distribution in the non-SC slabs.
In our simulations, we fix the thickness of the non-superconducting material and compute all of the eigenstates, which are functions of $k_x$ and $z$ ($k_y=0$).
The singlet pairing distributions for the NM, TI, and SM heterostructures are shown in fig.~\ref{NM_pairing}, \ref{TI_pairing}, \ref{SM_pairing}~(c,d), respectively. 
%{\clr SM pairing distribution}

\begin{figure}[t!]
\begin{center}
\includegraphics[clip,width=0.98\columnwidth]{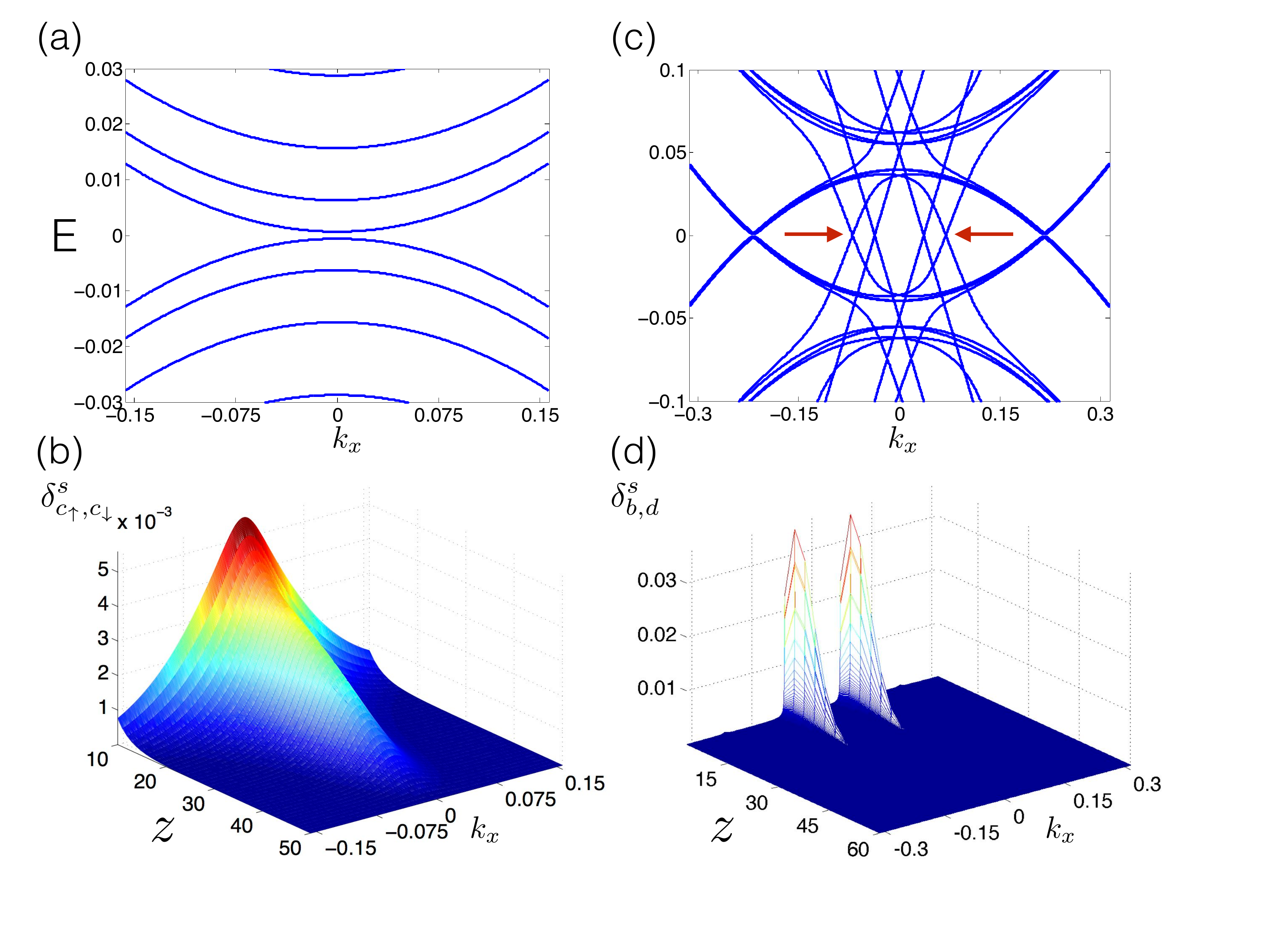}
\end{center}
\caption{To study gapped NM and TI slabs, we adjust the chemical potential $\mu_{\rm{NM}}=-2-\cos 0.05$ and $\mu_{\rm{TI}}=0.5$~eV. Panels (a) and (b) show the energy spectrum of the thin NM slab ($L_{\rm{SC}}=50$) and the pairing distribution $\delta_{c_\downarrow,c_\uparrow}^s$ as a function of $k_x$ and $z$ in the NM region respectively. The pairing decays without oscillation and the peak occurs at the smallest bulk gap of the NM. Likewise, panels (c) and (d) show the energy spectrum of the entire TI heterostructure ($L_{\rm{TI}}=60~\AA$) and the pairing distribution $\delta_{b,d}^s$ as a function of $k_x$ and $z$ in the TI region respectively. The coupling between the SC and TI layers is adjusted to $T=0.5$~eV since strong coupling dramatically changes the momentum of the surface states near the interface. The red arrows point out the $k_x$ location of the gapless surface states near the interface. This $k_x$ corresponds to the pairing peak.
}   
\label{decay_pairing} 
\end{figure}

Since spin $SU(2)$ symmetry is preserved in the NM heterostructure, $\delta_{c_\uparrow, c_\downarrow}^s(k_x,z)$ is the only pairing channel appearing and its spatial distribution is shown in fig.~\ref{NM_pairing}~(b,c).
To obtain a momentum independent measure of the pairing, we define a ``pairing average"
\bee
\overline{\delta^s_{c_\uparrow,c_\downarrow}}(z)=\sqrt{\sum_{k_x}|\delta^s_{c_{\uparrow},c_{\downarrow}}(k_x,z)|^2}
\ee
and fig.~\ref{NM_pairing}~(d) shows that  this average rapidly decays near the NM and SC interface and oscillates from one end to the other. On the other hand, for the TI heterostructure we consider an average over the two dominant singlet pairings  
\bee
\delta^s_{b,d}(k_x,z)=\sqrt{|\delta^s_{b_{\uparrow},b_{\downarrow}}|^2+|\delta^s_{d_{\uparrow},d_{\downarrow}}|^2},
\ee
and its $k_x$ independent pairing average
\bee
\overline{\delta^s_{b,d}}(z)=\sqrt{\sum_{k_x}|\delta^s_{b,d}(k_x,z)|^2},
\ee
The resulting spatial distributions are shown in fig.~\ref{TI_pairing}, where the singlet pairing similarly exhibits oscillations in the TI slab.
The pairing distribution when the non-SC slabs possess a bulk gap decays extremely rapidly away from the interfaces, shown in fig.~\ref{decay_pairing}.
We now separately consider the contributions to the singlet pairing distribution arising from SC bulk states, non-SC bulk states, and gapless surface states at the interface.

\begin{figure*}[ht!]
\begin{center}
\includegraphics[clip,width=1.5\columnwidth]{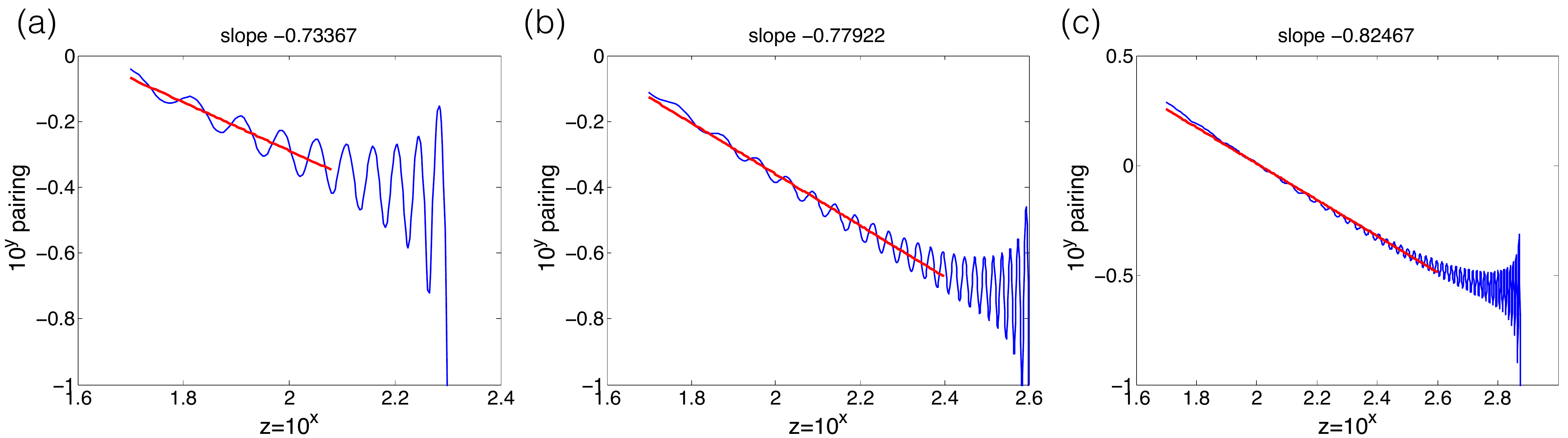}
\end{center}
  \caption{  The pairing distribution of the NM heterostructure: panel (a-c) show the the summation of the singlet pairing $\sum_{k_x}\delta^s_{b,d}(k_x,z)$ as a function of $z$ in log-log scale for $L_{\rm{NM}}=200,\ 400,\ 750$ respectively. The red fitting lines show the power (slope) of the decay rate is close to $-1$ as the system size is increased. }   
\label{one_over_z_decay} 
\end{figure*}

\paragraph{SC bulk states} The pairing arising from low-energy SC bulk states leaks in to the non-superconducting region. To confirm this SC bulk effect, we reduce the thickness of the NM slab so that the finite size effect gaps the bulk bands in the NM region, as described in fig.~\ref{decay_pairing} (a). This subfigure (b) shows that the singlet pairing decays from high to low near $z=0$ into the NM region without oscillations. The presence of SC bulk states is the only factor leading to this decay, which also explains the pairing decay from the interface to the NM region in fig.~\ref{NM_pairing}. However, the decay behavior completely vanishes for the TI heterostructure since the pairing is absent near the interface of SC and TI (see fig.~\ref{TI_pairing}~(d) near $z=0$). The main reason is the discrepancy between SC and TI Hamiltonians. Since SC and NM Hamiltonians in Eq.~\ref{HNM} are almost identical, except for the SC gap, the bulk SC wavefunction can easily leak into the NM to induce the pairing at the interface. However, due to the great discrepancy between SC and TI Hamiltonians in eq.~\ref{HFK}, the interface  becomes an effective open boundary so the pairing almost vanishes.

\paragraph{Non-SC bulk states} The structure of the bulk states in the non-SC region is responsible for the induced pairing oscillating spatially without decay.  Comparing with the other types of the induced pairings, this oscillating pairing dominates in the non-SC region. 
The oscillating pairing occurs at specific $k_{x}^n$, which correspond to zero-energy bulk states in the non-SC slab in fig.~\ref{NM_pairing}(a) and \ref{TI_pairing}(a).
The connection between $k_{x}^n$ and the number of pairing wave packets can again be simply explained by the wavefunctions $\phi_z^n$ (\ref{well}) in the 1D infinite square well in the $z$ direction.
%The wavefunction of the well in the $z$ direction can be written as 
%\bee
%\phi_z^n= \sin (k_z^n z ),
%\ee
%where $k_z^n=\pi n/L$ and $n$ is a positive integer. 
Since we are interested in zero energy, let us use the NM slab as an example so the momentum in the $x$ direction is determined by
\bee \label{kandn}
k_x^n=\cos^{-1}(-\cos(k_z^n)-1-\mu).
\ee
To have real solutions $k_x^n$, we have the constraint $|\cos k_z^n +1 + \mu|<1$.
Since $\mu=-2-\cos 0.1$ and $L=350$ in the stimulation, we have a maximum $n=11$ in agreement with the maximum number of the wave packets in fig.~\ref{NM_pairing}~(b).
Furthermore, the relation between $k_x^n$ and the number of the wave packets ($n\leq 11$) in fig.~\ref{TI_pairing} is exactly described by eq.~(\ref{kandn}). 

%need recover

\paragraph{Surface states}
The zero energy surface states located at the interface of the TI and SC also contribute to the exponential decay of the pairing in the TI region.
Since the gapless surface states are absent in the NM, we simply consider the TI heterostructure and adjust its chemical potential in the bulk gap region ($\mu=0.05$~eV) in fig.~\ref{decay_pairing}~(c,d); furthermore, the \emph{insulating} TI excludes bulk metallic non-SC inducing the pairing.  
%This setup excludes the two previous factors of bulk SC {\clr ???} and bulk non-SC states inducing the pairing.
By tracking the momentum location $k^0$ of the surface states at the interface we confirm the pairing decay occurs at that $k^0$.
As we redraw fig.~\ref{TI_pairing}~(d) in log-scale of pairing, the pairing exhibits oscillation, which stems from the wavefunctions of the surface states (see appendix \ref{surface derivation}). 
	 
It was believed that the induced pairing should always be weakened as the thickness of the non-SC slab increases. However, it is not the case as the slab transitions from bulk insulator to metal occurs. The reason is that the pairing is enhanced and supported across the entire slab by the presence of the zero-energy metallic non-SC bulk states.

%need recover

%
%It is quite surprising that the induced pairing is not weakened as the thickness of the non-SC slab is increased. {\clr ???}
%The reason is the pairing is supported across the entire slab by the presence of the zero-energy non-SC bulk states.

It is known that the momentum-independent singlet pairing exhibits $1/z$ decay\cite{Park_SC_DeGennes,PhysRev.132.1576}. Due to its momentum independence, we define the summation of the singlet pairing is  
\bee 
\tilde{\delta}^s_{\alpha,\beta}(z)=\sum_{k_x}\delta^s_{\alpha,\beta}(k_x,z)
\ee
However, if we simply compute singlet pairing $\tilde{\delta}^s_{c_\uparrow,c_\downarrow}(z)$ in the NM and metallic TI with the finite thickness, the pairing decay with additional oscillation is slower than $1/z$ since $1/z$ decay is only in the condition of the semi-infinite NM. Here, we consider the different sizes of the NM as shown in fig.~\ref{one_over_z_decay}. Although the power of the decay rate is never exactly $-1$, as the NM thickness is increased, the power is closer to $-1$. Furthermore, the oscillation along with the decay diminishes from the interface ($z=0$).

\subsection{Other pairing channels}

Triplet pairing, which breaks spin $SU(2)$ symmetry, is always absent in the NM heterostructure.
However, spin-orbital coupling in the TI and SM heterostructures (which breaks spin $SU(2)$ symmetry) mixes singlet and triplet pairing.
We might naively expect that induced triplet pairing in the heterostructures is much smaller than the induced singlet pairing from the proximity effect of the parent s-wave superconductor, but in the presence of strong spin-orbital interaction the two contributions can be comparable.

\begin{figure}[t!]
\begin{center}
\includegraphics[clip,width=0.98\columnwidth]{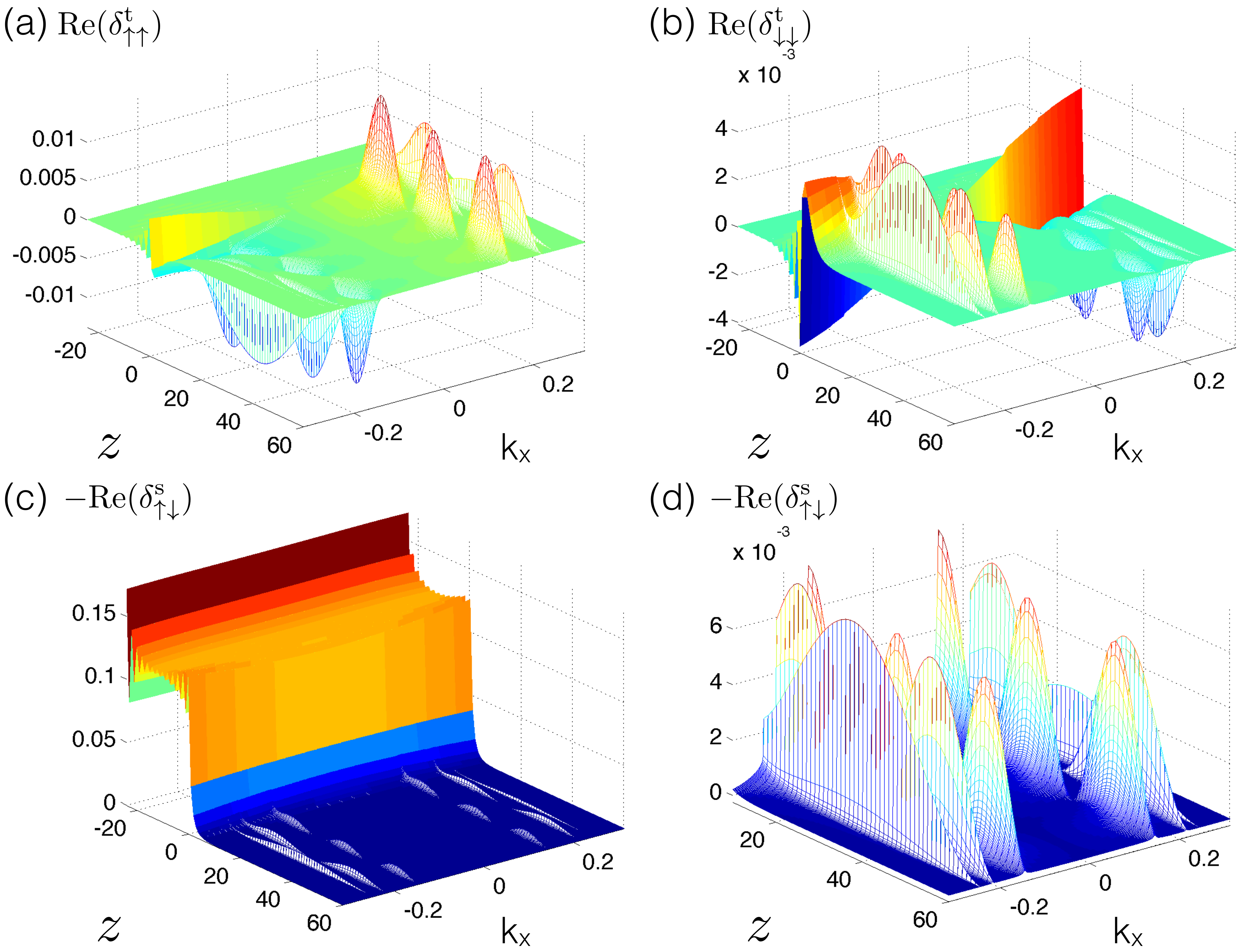}
\end{center}
  \caption{  The distribution of the pairing correlations in the SC ($-L_{\rm{SC}}<z\leq 0$) and SM ($0<z\leq L_{\rm{SM}}$) heterostructure as $L_{\rm{SM}}=60$. Non-zero pairing amplitudes appear at momenta corresponding to the SM Fermi level in fig.~\ref{heterostructure_TI_spectrum}~(d). (d) shows the semiconductor region only. 
}   
\label{SM_pairing} 
\end{figure}

%need recover

\begin{figure*}[t!]
\begin{center}
\includegraphics[clip,width=1.75\columnwidth]{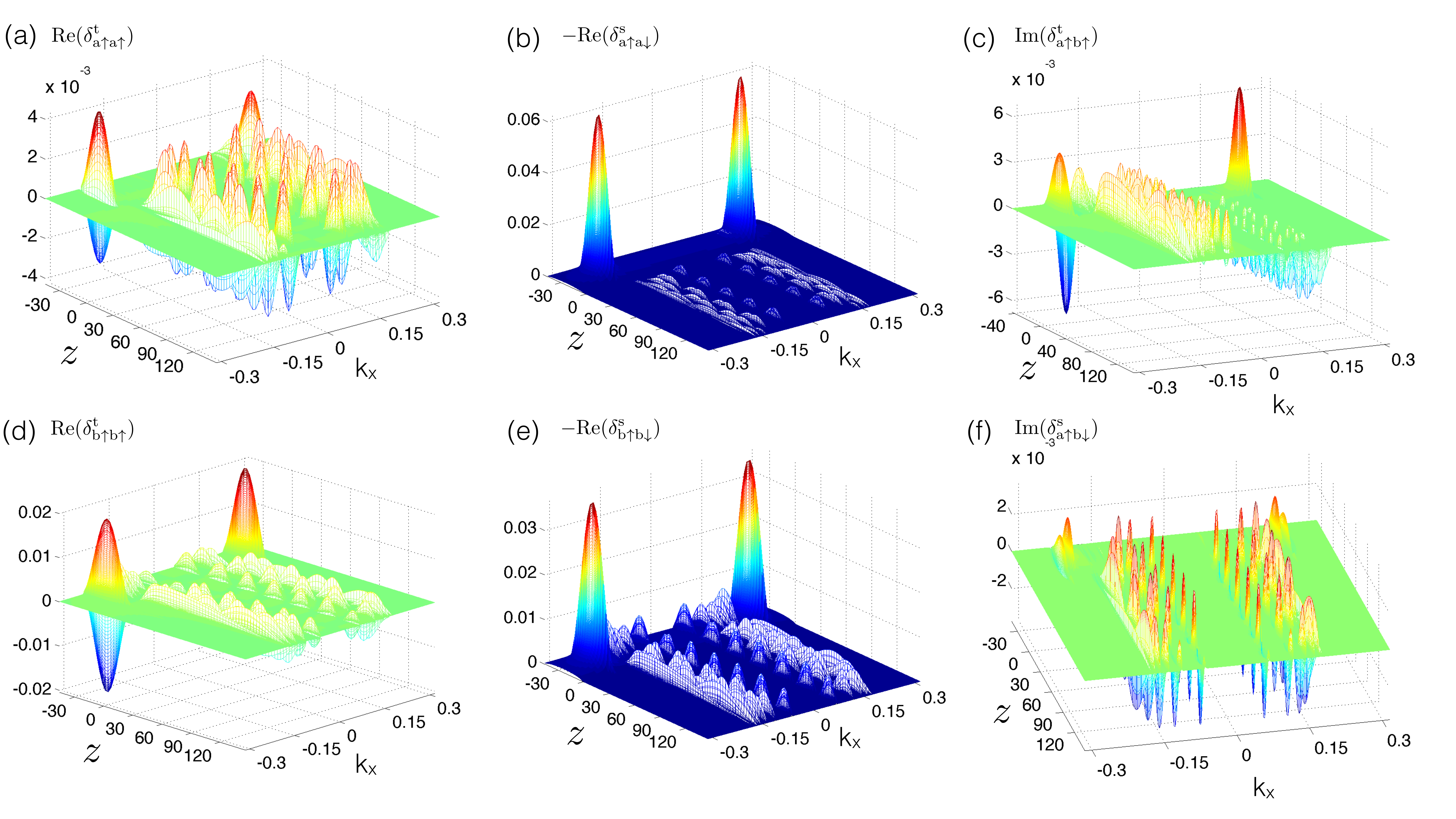}
\end{center}
\caption{Spatial distributions of non-zero singlet and triplet pairings in the SC ($-L_{\rm{SC}}<z\leq 0$) and TI ($0<z\leq L_{\rm{TI}}$) heterostructure. Panels (a), (c), (d) show odd functions of $k_x$ (triplet pairing) whereas panels (b), (e), (f) show even functions of $k_x$ (singlet pairing). While panels (b) and (e) demonstrate that the singlet pairing dominates in the SC region, the triplet pairing in panel (d) has a comparable weight distribution to the singlet pairings in the TI region.
}   
\label{All_pairing} 
\end{figure*}

Let us first discuss the SM heterostructure. The Hamiltonian $\hat{H}^{\rm{SM}}$ (\ref{HSM}), which includes only the spin degree of freedom, can be rewritten in the particle-hole basis
\bee
H^{\rm{SM}}_z(\bk)\equiv
\bma
H_o(\bk) & H_\Delta(\bk) \\
H_\Delta^\dagger(\bk) & -H_o(-\bk)
\ema.
\ee
By solving the eigenvalue problem of $\hat{H}^{\rm{SM}}$, from the eigenstates we can obtain the projection matrix 
\begin{align}
P^{\rm{SM}}_z(\bk)=&\sum_{E\leq 0} \ket{\Psi_E(\bk, z)}\bra{\Psi_E(\bk, z)} \nonumber \\
\equiv&
\bma 
P_o^{\rm{SM}}(\bk) & P_\delta^{\rm{SM}}(\bk) \\
{P_\delta^{\rm{SM}}}^\dagger(\bk) & -{P_o^{\rm{SM}}}^*(-\bk)
\ema.
\end{align}
The off-diagonal term of the projection matrix determines  all of the singlet and triplet pairings
\bee
P_\delta ^{\rm{SM}}(\bk )= 
\bma 
\delta^t_{\uparrow \uparrow } & \delta_{\uparrow \downarrow}^t + \delta^s_{\uparrow\downarrow} \\
\delta_{\uparrow\downarrow}^t - \delta_{\uparrow\downarrow}^s & \delta_{\downarrow \downarrow}^t
\ema.
\ee	
We study the distribution of the pairings in the momentum space as $L_{\rm{SM}}=60$ form the projection matrix of $\hat{H}$. Fig.~\ref{SM_pairing} shows non-zero pairing amplitudes in the semiconductor corresponding to the zero energy momenta (c.f. the spectrum in fig.~\ref{heterostructure_TI_spectrum}~(d)). Furthermore, the singlet pairing dominates in the SC region as shown in fig.~\ref{SM_pairing}~(d) due to the original s-wave superconductivity. Only the presence of the spin orbital coupling gives rise to the triplet pairings. As the spin orbital coupling is turned off, the Zeeman effect by itself is unable to induce triplet pairing. Instead, in the presence of spin-orbital coupling, the Zeeman effect $F=-0.01$ leads to a situation where the strength of the up-up pairing is greater than the down-down pairing.

The superconducting pairing channels in the TI heterostructure include two spin and two orbital degrees of freedom; hence, there are 16 types of triple and singlet pairings.
Since the heterostructure preserves time reversal symmetry ($\tau_0s_y H^{\rm{TI*}}(-\bk)\tau_0s_y= H^{\rm{TI}}(\bk)$), we can use this to further reduce the number of the non-zero independent pairings. Similarly, the projection matrix is determined by the occupied states in the TI heterostructure. 
%To preserve TRS, the Hamiltonian obeys $\tau_0s_y H_{\rm{SC}}^*(-\bk)\tau_0s_y= H_{\rm{SC}}(\bk)$.
\begin{widetext}
The $4\times 4$ off-diagonal matrix $P_\delta(\bk)$ contains all 16 pairings channels.
Additionally, though, the projection matrix inherits TRS from the BdG Hamiltonian.
After applying these symmetry constraints, we can write the pairing matrix in a reduced, generic form 
\bee
P_\delta (\bk)=
\bma 
\delta^{t}_{a\uparrow a \uparrow} & \delta^{t}_{a\uparrow a \downarrow}+ \delta^{s}_{a\uparrow a \downarrow} & \delta^{t}_{a\uparrow b \uparrow}+ \delta^{s}_{a\uparrow b \uparrow} 
& \delta^{t}_{a\uparrow b \downarrow}+ \delta^{s}_{a\uparrow b \downarrow}  \\
\delta^{t}_{a\uparrow a \downarrow} - \delta^{s}_{a\uparrow a \downarrow} & -\delta^{t*}_{a\uparrow a \uparrow} 
& \delta^{t}_{a\uparrow b \downarrow} - \delta^{s}_{a\uparrow b \downarrow} &  -\delta^{t}_{a\uparrow b \uparrow}+ \delta^{s}_{a\uparrow b \uparrow}  \\
\delta^{t}_{a\uparrow b \uparrow} - \delta^{s}_{a\uparrow b \uparrow} 
& \delta^{t}_{a\uparrow b \downarrow}+ \delta^{s}_{a\uparrow b \downarrow} & \delta^{t}_{b\uparrow b \uparrow} & \delta^{t}_{b\uparrow b \downarrow}+ \delta^{s}_{b\uparrow b \downarrow}    \\
\delta^{t}_{a\uparrow b \downarrow} - \delta^{s}_{a\uparrow b \downarrow} & -\delta^{t}_{a\uparrow b \uparrow} - \delta^{s}_{a\uparrow b \uparrow} 
   & \delta^{t}_{b\uparrow b \downarrow}- \delta^{s}_{b\uparrow b \downarrow} & -\delta^{t*}_{b\uparrow b \uparrow}   \\
\ema (\bk). \label{constraint}
\ee
The constraints from the symmetries additionally limit $\delta_{a\uparrow a\uparrow}^{ t},\delta_{b\uparrow b\uparrow}^{ t}\in\mathcal{R}$.
The symmetries simplify the superconducting pairings down to 10 undetermined parameters $\delta^t_{a\uparrow a \uparrow},$$\ \delta^t_{a\uparrow a\downarrow},$ $\delta_{a\uparrow a\downarrow}^s$, $\delta^t_{a\uparrow b\uparrow},$ $\delta^s_{a\uparrow b\uparrow},\ \delta^t_{a\uparrow b\downarrow},\ \delta^s_{a\uparrow b\downarrow},\ \delta^t_{b\uparrow b\uparrow},\ \delta^t_{b\uparrow b\downarrow},\ \delta^s_{b \uparrow b\downarrow}$, which are then dependent on specific details of the model. 
\end{widetext}

Now we can consider all the remaining possible pairings in the TI heterostructure.
Since the pairing obtained from unbiased numerics satisfies these symmetry constraints, this also provides a consistency check on our calculations.
Moreover, in this TI heterostructure $\delta^t_{a\uparrow a\downarrow}, \delta^s_{a\uparrow b\uparrow}, \delta^t_{a\uparrow b\downarrow}$, and $\delta^t_{b\uparrow b\downarrow}$ are always zero.
The spatial distributions of the six non-zero pairings are shown in fig.~\ref{All_pairing}. Inside the SC region, singlet pairing dominates since the singlet pair potential $H_\Delta=\Delta i\sigma_y$ is included in the model by hand.
In the TI region, triplet and singlet pairings oscillate along the $z$ direction with almost comparable weight distributions.

\section{Summary and Conclusions} \label{summary}

	The SC proximity effect induces two independent quantities -- spectral gap and pairing correlations -- in the non-SC slab due to the absence of intrinsic electron-phonon interaction. Although the quality of the electrical contact between SC and non-SC does significantly affect the induced gap, as the thickness ($L$) of the non-SC slab is varied, the decay behavior of the gap on the naked surface should only depend on $L$.
	The decay behavior is characterized for three different regimes of the non-SC slabs -- insulator, metal, and topological insulator. When the non-SC slab is an insulator, since the SC gap is small, the induced gap can be neglected compared to the bulk gap of the insulator. When the non-SC slab is a generic metal, the induced spectral gap exhibits $L^{-3}$ decay. Once the non-SC slab becomes a topological insulator, the induced spectral gap, which originates from the gapless states on the naked surface leaking into the SC region, exhibits exponential decay $e^{-L/\xi}$. However, the decay behavior of the gap leads to a dilemma for MZM realization in the TI/SC heterostructure. If the topological insulator is bulk-insulating, the strength of the induced gap is close to the strength of the MZM hybridization between both TI surfaces since these two quantities are determined by the spatial distribution of the MZM wavefunction, this creating an undesirable situation where the Majorana splitting is comparable to the induced gap. On the other hand, if the topological insulator is bulk-metallic, then low-energy Caroli-de Gennes-Matricon modes\cite{CAROLI1964307} are present in the vortex, and MZMs are indistinguishable from the usual low-energy intra-vortex bound states in a superconductor. Hence, to realize MZMs in the TI heterostructure requires fine tuning to produce a strong induced gap and avoid the presence of extra low-energy vortex modes.  Perhaps intrinsic superconducting TIs\cite{SC_SrBiSe,2015arXiv151106231C,2015arXiv151106942X,Zhao:2015aa} might be an alternative direction for realizing TSCs.

	The oscillating induced pairing in the non-SC occurs at momenta corresponding to zero energy in the non-SC spectrum. That is, the presence of this pairing requires that the non-SC slab is a metal. Furthermore, the summation of the pairing in momentum space exhibits $1/z$ decay inside the non-SC slab, which is consistent with prior results. When the entire heterostructure preserves spin $SU(2)$ symmetry, any triplet pairing should be absent. When the non-SC slab includes spin orbital coupling, the triplet pairing can be induced in the entire heterostructure. In the non-SC region, the triplet and singlet pairings may have equal strength, despite arising from proximity with an s-wave SC.

	In short, the induced spectral gap exhibits sharper decay than the induced pairing. These distinct decay behaviors are consistent with the observed rapid decay rate of the induced gap\cite{SC_Proximity_Jia,Xu_TI_SC} and the slow decay rate of the pairing\cite{Proximity_Josephson} seen in experiments of NM\cite{PhysRevLett.77.3025} and metallic TI\cite{Proximity_Josephson}.  Our most important fundamental insight provided in the current work is explaining with concrete results the clear conceptual distinction between pairing and gap in the superconducting proximity effect involving non-superconducting materials, which is of great significance in the current search for topological superconductivity and Majorana fermions in various heterostructure systems involving superconductors and topological insulators or semiconductors.
	
\section{Acknowledgements}

The authors are indebted to A. J. Leggett, J. D. Sau, and T. D. Stanescu for discussions. 
This work is supported by Microsoft Q and LPS-MPO-CMTC.

%\begin{enumerate}
%
%\item The induced SC gap exhibits $L_{\rm{SM}}^{-3}$ decay as the thickness $(L_{\rm{SM}})$of the semiconductor is increased.  
%
%\item The triplet pairing amplitudes in semiconductors are induced by the spin orbital coupling, which breaks spin $SU(2)$ symmetry. The Zeeman field in the $z$ direction does not induce any types of pairing. Instead, as the triplet pairings are present, the Zeeman field leads to the imbalance of the up-up and down-down triplet pairings. 
%
%\end{enumerate}

%------------------------------------------------------------------------------%

\appendix

\section{Analytic solution of the TI surface states}\label{surface derivation}
The BdG Hamiltonian $\hat{H}_{\rm{TI}}$ (\ref{TI_FK}) under a proper basis transformation can be simply decomposed into 4 copies of 1D $2\times 2$ tight-binding models with open boundary conditions
\begin{align}
\hat{H}_\pm =&\pm \sum_{z=a}^{L}C^\dagger_z \big(\nu \sigma_x + \eta \sigma_0\big)C_z  \nonumber \\
\pm&  \big (\sum_{z=2a}^{L-a} C^\dagger_{z}  \frac{\alpha\sigma_x+i\beta\sigma_y}{2} C_{z-a}+ h.c. \big ),
\end{align}
where $C_z=(A_z\ B_z)$, $\eta=A_2 \sin k_x -\mu$, $\nu=M_{\rm{SC}}- 2B_1 - 2B_2 (2- \cos k_x )$, $\alpha=2B_1$, and $\beta=A_1$. The signs $\pm$ correspond to PHS eigenvalues. We note that with periodic boundary conditions the Hamiltonian in momentum space is given by 
\begin{align}
H_\pm(k_z)=\pm \big (  \eta\sigma_0 +(\nu + \alpha \cos k_z ) \sigma_x + \beta \sin k_z \sigma_y   \big )
\end{align}
When $|\nu|<|\alpha|$, this model hosts gapless surface states. Back to the chain with open boundary conditions, to find the surface states we write  
the eigenstate in the form of
\bee
\ket{\Psi}=\sum_{z=0}^L (a_z A^\dagger_z + b_z B^\dagger_z ) \ket{\bf 0},
\ee
and we are interested in the wavefunction at zero energy 
\bee 
0=\hat{H}_\pm \ket{\Psi}. 
\ee
We note the solutions are independent of $\pm$ so the recursion relations are given by 
\begin{align}
\frac{\alpha + \beta}{2}a_{z+a} +\nu a_z + \frac{\alpha-\beta}{2} a_{z-a} =&0, \\
\frac{\alpha - \beta}{2}b_{z+a} +\nu b_z + \frac{\alpha+\beta}{2} b_{z-a} =&0,
\end{align}
here we adjust $k_x$ so $\eta=0$. The open boundary conditions lead to 
\bee
a_{0}, b_{0}, a_{L+a}, b_{L+a}=0.
\ee
Let us focus on the $z=L$ surface, since we are interested in the distribution of this surface state near $z=0$, which is related to the induced superconducting gap. By solving the recursion relation with the boundary condition, we obtain two independent solutions 
\begin{align}
a_z=&N_a(\gamma_+^{z/a-l-1}-\gamma_-^{z/a-l-1}),             \\
b_z=&N_b(\gamma_+^{-z/a+l+1}-\gamma_-^{-z/a+l+1}),
\end{align}
where $l=L/a$ and $N_{a,b}$ are normalization factors and
\bee
\gamma_\pm = \frac{-\nu\pm \sqrt{-\alpha^2+\beta^2 +\nu^2}}{\alpha+\beta}.
\ee
To simplify the problem, we assume $\alpha,\ \beta$ are positive. Since we expect the surface states are localized at the right end, to have normalizable wavefunctions $a_z$ or $b_z$ goes to zero as $n\rightarrow -\infty$. That is, $|\gamma_\pm| >  1$ for $a_z$ and $|\gamma_\pm| < 1$ for $b_z$. First, consider $\alpha^2< \beta^2 +\nu^2$. As $|\nu|<|\alpha|$, which is the topological region, 
$|\gamma_\pm|<1$ leads to the only normalized surface state $\sum_z^{L}b_z\beta^\dagger_z \ket{\bf 0}$. Hence, this wavefunction at $z=\epsilon$ close to $0$ exhibits exponentially decays as $l$ is increased. However, as $|\nu|>|\alpha|$, the absolute value of one of $\gamma_\pm$ is greater than 1 and the other is less than 1 so these two candidates of the surface states cannot be normalized in agreement with the region of trivial topology. Second, let us consider the other region $\alpha^2> \beta^2 +\nu^2$ so we can rewrite 
\bee
\gamma_\pm =\sqrt{\frac{\alpha-\beta}{\alpha+\beta}}e^{\pm i\theta},
\ee
where $\theta=\rm{angle}(-\nu+i \sqrt{\alpha^2-\beta^2-\nu^2})$, which might be any number. Since $\alpha^2> \beta^2 +\nu^2$ and $\alpha, \beta>0$ lead to $|(\alpha-\beta)/(\alpha+\beta)|<1$, $\sum_z^{L}b_z\beta^\dagger_z \ket{\bf 0}$ is the only normalizable surface wavefunction. The portion of the surface state leaking to the SC region is given by 
\begin{align}
|\Psi_{\rm{surf}}^{\rm{leak}}| \sim b_\epsilon=2i N_b\big (\frac{\alpha-\beta}{\alpha+\beta} \big )^{\frac{-\epsilon+l+1}{2}} \sin \big ((-\epsilon+l+1)\theta \big ) .
\end{align}
Due to the presence of ``$\sin$'', as the thickness of the TI slab $l$ is increased, the wavefunction at $z=\epsilon a$ exponentially decays with oscillation. In the Bi$_2$Se$_3$ heterostructure, $\alpha^2> \beta^2 + \nu^2$ so the induced superconducting gap also exhibits oscillating exponential decay. Hence, as the TI is a true insulator, we can use $b_\epsilon$ to fit the induced gap 
\bee
\Delta_{\rm{I}} = \zeta \big  |\frac{\alpha-\beta}{\alpha+\beta} \big |^{-\epsilon+l+1} \sin^2 \big ((-\epsilon+l+1)\theta \big ),
\ee
where $\zeta, \epsilon$ are determined from the fitting. Fig.~\ref{small_L} (c) shows the surface state contribution provides an accurate fit to the simulation. Interestingly, the exponential decay rate is always a constant, which is independent of the chemical potential. On the other hand, the period of the oscillation strongly depends on the chemical potential. Another interesting fact is that this oscillation also can explain the energy oscillation of the domain wall moving along the 1D Kitaev chain\cite{Chiu_oneD_Majorana_move}.

\bibliographystyle{apsrev4-1}
\bibliography{TOPO}

\end{document}